# Harmonizing Signals and Events with a Lightweight Extension to Java

Tetsuo Kamina[a] and Tomoyuki Aotani[b]

a Ritsumeikan University, Japan

b Tokyo Institute of Technology, Japan

**Abstract** The current demands for seamless connections with the surrounding environment make software more reactive. For example, such demands are evident in systems consisting of the Internet of Things. Such systems include a set of reactive values that are periodically updated in response to external or internal events to form a dataflow in the sense that such updates are propagated to other reactive values. Two effective approaches for realizing such reactive values have been proposed: the event mechanisms in event-based programming and the signals in functional-reactive programming. These two approaches are now becoming mixed in several languages such as Flapjax and REScala, which makes these languages notably expressive for modularizing the implementation of reactive software. For example, REScala provides a rich API that consists of functions converting events to signals and vice versa.

In this paper, we explore another, simpler approach in the design space of reactive programming languages: the event mechanism is harmonized with signals, resulting in a simplified programming interface that is mostly based on signals. Based on this approach, we realize SignalJ, a simple extension of Java with events and signals. Our notable findings are (1) an event can be represented as an update of a signal and (2) such an effectful signal can be represented using annotations instead of introducing types and constructors for signals to further simplify the language.

Another contribution of this paper is the formal model of SignalJ. As both mechanisms of events and signals may interfere with each other, this mixing sometimes results in surprising behavior. For example, the functional behavior of signals is affected by the imperative features of events. Thus, understanding the formal model of this mixing is actually important. The core calculus, Featherweight SignalJ (FSJ), was developed as an extension of Featherweight Java, and proofs are provided to ensure the soundness of FSJ.

## 1 Introduction

The current trend of seamless connections between computing systems and their surrounding environments, such as cyber physical systems and the Internet of Things, is making reactive software increasingly important. Such systems include a set of *reactive values*, i.e., values that are periodically updated in response to external or internal events. These reactive values also form a dataflow, where an update of a reactive value is propagated to other reactive values. Traditionally, such reactions are implicitly implemented using coding patterns cluttered with protocols that are unrelated to the application logic itself, which makes the implementation of reactive systems challenging.

Besides such traditional approaches, several approaches have been used to naturally implement reactive values. First, event-based languages that support first-class events at the language level [32, 13] have been proposed. They support modularity and are integrated with object-oriented (OO) design, even though they generate events using imperative operations, which make it difficult to determine data dependencies with respect to the reactive values [36]. Second, there have been languages that provide constructs for networks of time-varying values, as in synchronous dataflow languages, such as Esterel [4] and Lustre [17], as well as functional-reactive programming (FRP) languages [12, 38, 7, 28]. Data dependencies are declaratively and naturally represented in dataflow and FRP languages. Even though existing FRP languages do not fit well into the imperative setting, they would compensate for the weaknesses of event-based languages by using declarative specifications and automatic updates between reactive values. There are also other reactive programming languages such as reactive extensions (Rx), which only provide event streams, and Flapjax [28] and REScala [33] that provide both signals and events at the syntactic cost of introducing conversion functions between them.

In this paper, we propose SignalJ, a simple extension of Java that harmonizes events and signals in that the event mechanism is realized in terms of signals. The novelty of SignalJ is its way of combining imperative events with FRP signals. First, instead of providing interfaces for both events and signals, SignalJ provides the interface for only signals, while events are just signal updates and handlers on signals are used to provide side effects. Second, instead of introducing types for signals, a signal is annotated using a modifier. This annotation approach further simplifies the language semantics, as the conversions between non-signal values and signals in existing languages are no longer necessary.

As this harmonization raises the necessity of its theoretical basis, we developed Featherweight SignalJ (FSJ), a core language of SignalJ that is based on Featherweight Java (FJ) [19]. This is because signals and events may interfere with each other, which makes the program behavior unpredictable without precise knowledge of change evaluation strategies and update ordering. For example, we adopt the *pull-based* evaluation strategy for signal update (i.e., a signal is reevaluated only when it is accessed) where the update order is determined according to the dataflow, while the event mechanism is inherently *push-based* (event-driven). This mixing of evaluation strategies motivates us to develop a formal calculus of SignalJ. FSJ carefully determines the order of execution, and a set of theorems regarding the type safety ensures the soundness of this calculus.

Our contributions are summarized as follows:

- The design of an object-oriented programming language, SignalJ, that harmonizes push-based events with pull-based signals. The programming interface is mostly simplified to be signal-based.
- A type safety proof for the core calculus of SignalJ.

The rest of this paper is structured as follows. Section 2 reviews the concepts of events and signals and explains our motivation. Section 3 presents SignalJ as a simple language of event-signal unification. Section 4 formalizes the core calculus of SignalJ. Finally, Section 5 discusses the related work.

## 2 Events, Signals, and Their Harmonization

Event-based programming has been proposed to address the limitations of procedural languages in implementing reactive behavior. These limitations include the fact that procedural languages heavily depend on callbacks to represent reactivity, where the inversion of control makes the program difficult to understand, analyze, and achieve separation of concerns. Event-based programming languages provide linguistic constructs for representing events and event-driven interactions. These languages include a mainstream language such as C#, which provides *event declarations* where a declared event is broadcasted to all preregistered event handlers. More advanced features have also been studied as an extension of existing languages such as EventJava [13], Ptolemy [32], and EScala [16].

Even though event-based languages provide abstractions for implementing sophisticated event-driven behavior such as event composition and correlation, they also have their own drawbacks in implementing reactive behavior. For example, event generation is basically an imperative operation, and the response to the event is represented using an *event handler*, which makes it difficult to determine the dependencies among data and behavior from the source code. This drawback is explained using motor control software as an example, where the power of the motor is calculated according to the current sensor value. In an event-based language, a change in the sensor is represented as an event. For example, in EScala [16], this is achieved by calling an event, namely, changed, assuming that the object sensorValue has that event:

```
sensorValue.changed(newValue);
```

The behavior executed when that event is fired is registered as an event handler that calculates the power difference for the motor as follows:

```
sensorValue.changed += { (e) -> powerDifference = f(e) }
```

Whereas the firing of an event is placed in the code that reads the sensor value, the event handler is declaratively represented. Thus, both are separately represented, and the dependency between sensorValue and powerDifference is implicit. Even though event handlers are useful for describing reactions to effects [3], the implementation of the functional dependencies among data and behavior using an event handler is tedious and error-prone.

It is easy to represent functional dependencies in functional-reactive programming (FRP). There are several FRP languages and directives such as Fran [12], FrTime [7], and Flapjax [28]. These languages handle *time-varying values* and the functions over them. For example, consider the following equation that represents the relationship between the sensor value and the power difference using the function f:

```
powerDifference = f(sensorValue)
```

In FRP languages, sensorValue is a function over time. Thus, by accessing sensorValue, we can obtain the value of the sensor *at this time*, and by accessing powerDifference, the power difference *at this time* is obtained accordingly. The function f is considered functional, i.e., it should not produce side effects. Thus, unlike event-based languages, we can directly represent the functional dependency between the sensor and the motor.

### 2.1 Design Space for Mixing Events and Signals

The FRP mechanism, also known as a signal, was recently realized in an imperative OO setting where the imperative update of signals is allowed. For example, REScala integrates signals with the event mechanism. This appears to be convenient for programmers, as it is notably expressive for modularizing the implementation of reactive software [33]. This is also useful when signals interact with existing imperative programs. For example, the motor will be accompanied by the API for controlling it, which is implemented as a legacy library where the power of the motor is not represented as a signal. In such a case, the use of an event handler is a convenient way to propagate the changes in the signal to the actuator as follows

```
powerDifference.changed += { (e) -> adjustMotor(e) }
```

where changed is the REScala built-in function that returns an event that is fired everytime the value of the signal is changed, and adjustMotor is a library function for changing the power of the motor.

The approach of REScala to mix events and signals is using conversion functions between them. REScala provides a rich API that consists of functions converting events to signals and vice versa. Conversion functions from signals to events seamlessly integrate FRP-like computations into event-driven ones. One of the basic conversion functions from signals to events is changed, which was explained above. Providing more sophisticated conversion of signals to events, REScala supports a variety of representations to integrate signals with event-based computations.[1]

Although this approach would mostly have enough power to represent both event-based and signal-based computations, we can still explore another, simpler approach in the design space of reactive programming languages: events can be considered side effects of signal update, resulting in a simplified programming interface that is mostly based on signals. This interface is easy to be integrated with the existing industry standard language, such as Java, that supports neither events and signals. Furthermore, the semantics for events and signals are nicely combined by equating event firing with signal update. For example, we can formally define the ordering of update propagation and event firing (Section 2.2) without explicitly converting events to signals.

[1] REScala also supports conversions from events to signals, to make event-based imperative computations can be wrapped up and abstracted over in signal-based functional computations.

### 2.2 Differences in Propagation Strategies

Another issue that we address in this paper is how to *harmonize* the different evaluation strategies that are taken by signal-based computations and event-based ones. Signals represent behavior that iteratively executes over time-varying values [35], which are sequences of sampled values from sensors that monitor the surrounding environment. This means that time does not proceed within each iteration. Instead, each iteration runs at a different time. In the following PI control system, for example, the signals sum and diff are *immediately* changed when sensorValue is updated (we continue using REScala and its conversion functions to explain the issue):

```
val sum : Signal[Int] = sensorValue.changed.fold(0)(_ + _)
val diff : Signal[Int] = Signal{ P*sensorValue() + I*sum() }
```

In this example, the proportional value is the value of the sensor itself, and the integral value is obtained as the sum of the past proportional values. The signal sum holds the sum of all elements in the event stream sensorValue.changed. P and I are preset parameters for PI control that determine how the proportional and integral values affect the power of motor, respectively. If the value of sensorValue is obtained at the time $t$, it is ensured that the values of the other signals, sum and diff, are also obtained at the time of $t$.

The implementation for this immediate update is an issue. One approach is *push-based*, where changes in sensorValue are eagerly propagated to all sinks. In this approach, we need to consider the order of propagation. For example, Flapjax [28] provides topological sorting to ensure that sum is reevaluated before the reevaluation of diff. Another approach, which we adopt because of its simplicity, is *pull-based*, where reevaluation only occurs when the signal is used. For example, diff is reevaluated at the time it is accessed using the current value of sum, i.e., the value of sum is reevaluated before diff is reevaluated. Thus, the evaluation order is determined according to the dataflow, which is the same as Flapjax's topological sorting.

On the other hand, event-based programming represents its reactivity by invoking handlers in response to events that are fired during the execution of the program. In this setting, time proceeds when an event is fired. Using the event mechanism, we extend the above PI control system to the following PID control system and implement the propagation to the actuator[2]:

```
var last : Int = /* some initial value depending on the application */
sensorValue.changed += { x => last = x }
val diff : Signal[Int] =
  Signal{ P*sensorValue() + I*sum() + D*(sensorValue() - last) }
diff.changed += { x => adjustMotor(x) }
```

The derivative value is obtained by subtracting the last proportional value last, which is calculated by the event handler registered to the changed event of sensorValue, from the current one. We note that the event handlers are called in the push-based manner, as the wording "event-driven" suggests; that is, each handler is eagerly called when any change in the corresponding signal is observed.[3]

Then, one question arises: what is returned if the value updated by the event handler is accessed within the signal? For example, in the above example, last is updated by the event handler and accessed within the signal diff. To make this program work appropriately, the event handler should be called *after* update propagation, which ensures that last retains the value of sensorValue before it is updated.

This question becomes even more difficult when the subscription of an event handler occurs *during the signal updates*. The following code fragment shows an example:

```
val diff : Signal[Int] = ... /* same as above */
calibrate(sensorValue)
diff.changed += { x => adjustMotor(x) }
```

This example first defines the signal network in the same manner as the above example; then, it calibrates the sensor, and the event handler is registered to react to the changes in the calculated power difference. At first, FRP programmers would wonder if the event handler would react to the updates during calibration because the signal should contain all of the sampled values. Actually, the event handler in REScala reacts to the updates only after the imperative subscription. Thus, when mixed with events, the functional behavior of a signal is affected by the imperative features of events.

However, this example shows only an execution instance. There is no clear language semantics that explains why this example behaves in this manner.

## 3 SignalJ by Examples

The above discussions lead us to a design of new language, SignalJ, which is a lightweight extension to Java with signals where the event mechanism is supported in terms of signals, and useful to formalize the language semantics.

[2] Even though the last sensor value could be obtained using the fold function, we introduce the mutable state last to explain the differences in signal and event propagations.

[3] Even though we may also consider event handlers that are called *on-demand* (e.g., as in Adapton [18]), we consider that most event-based systems call event handlers eagerly. In this paper, we focus on such event-based systems.

### 3.1 Static Signal Networks

In SignalJ, we can declare signals, which are used to represent functional dependencies between values that are updated during computation. This dependency is specified in a declarative way. A signal is declared using the signal modifier. For example, the following code fragment declares two signals, a and b, where b depends on a:

```
signal int a = 5;
signal int b = a + 3;
a++;
System.out.println(b); // display 9
```

We refer to a signal that depends on other signals as a *composite signal*, which represents a functional dependency between signals. In SignalJ, a signal depends on all signals that appear on the right-hand side of the initialization (=) of the signal. For example, b in the above code fragment is a composite signal that depends on a; that is, when the value of a is updated, this update is propagated to b, resulting in the update of the value of b. This dependency is fixed during execution. This means that the reassignment of a value to b is not allowed in SignalJ, and the value of b is updated only through the update of a.

On the other hand, a in the above code fragment does not depend on any other signal. It is considered a source of the signal network, and its value can be *imperatively* updated at runtime. We refer to this signal as a *source signal*. The change in the source signal is implicitly propagated to other signals that depend on this source signal. Thus, the value of b in the above code fragment is initially 8, but after the value of a is updated by a++, the value of b becomes 9. The updates of all of the dependent signals are performed at the time of the update of the source signal. Thus, the update of b simultaneously occurs when a is updated, and 9 is displayed when println is called after a++. We note that a source signal must be provided an initial value, as every signal needs to have an initial value.

### 3.2 Annotation Approach for Distinguishing Signals

A notable feature of SignalJ is its way of distinguishing signals with other non-reactive values. Instead of introducing types and constructors for signals, SignalJ distinguishes a signal using annotations (i.e., the signal modifier). This means that signal int and int are the same type.[4] Thus, a signal can be used in the place where a non-signal is expected, and it is evaluated to the current value in such a case. Because any conversions between signals and non-signals (also known as lifting and unlifting) are not necessary, this approach simplifies the language semantics. To our knowledge, this is the first type system that signal, which differentiates signals from non-signals, actually does not differentiate values (Section 4).

[4] Precisely, signal is more than a modifier, as it allows the declared variable to accept some specific operators, such as subscribe, which are not allowed to call on non-signals.

From the viewpoint of programmers, this annotation approach is especially useful for creating a network of signals using legacy Java libraries. We take an example that counts the number of one-bits in the two's complement representation of a given signal:

```
signal int a = 0; // updated later
signal int count = Integer.bitCount(a);
```

The method bitCount is provided by the Java standard library. It takes an int value as an argument and returns an int value. Owing to the annotation approach, signal a can be an argument to bitCount, and the return value is used to initialize count, which is a signal. As count is declared as a signal, when a is updated, bitCount is automatically recomputed, and count is updated[5]

One may criticize that this approach may disturb program comprehension. For example, when we insert/remove signal to/from a field declaration, the compiler will not report errors, but the program behavior may be changed drastically. We are not sure whether such modification is realistic (whether a field is made a signal may be determined at the time of the design). At least, our mechanism can provide an aid to program comprehension. For example, it is easy to equip a highlighting mechanism that distinguishes signals from non-signals with an editor, as the type system actually knows which variables are signals (Section 4.3).

### 3.3 Dynamic Signal Networks

The construction of signal networks can also be dynamic. This is due to the fact that we can construct a signal of an object that encapsulates other signals.

For example, we assume a multi-tabbed content view that is updated at runtime. The current view is designed as a signal because its value is updated at runtime. Furthermore, this current view also depends on the selected tab; that is, there is a set of views, and which view is selected is also represented as a signal. This motivates us to switch the connection of signal networks dynamically.

The object-typed signals can naturally realize such a dynamic construction of signal networks. First, a multi-tabbed view is modeled as a list of views, where each view includes content that is declared as a signal, which is declared as a String for simplicity:

```
class View {
  signal String content = ...;
  View() { content.subscribe(this::drawString); }
}
List tabs = new List().add(new Tab(new View()));
```

The dependency between the selected tab and the current view is then represented by declaring each of them as a signal:

[5] The legacy code does not have to be recompiled. The compiler translates each signal use into a method call on that signal that returns the current value of the signal. Each field initializer (for a composite signal) is translated into a lambda expression that is evaluated when that signal is used (Section 3.7).

```
signal int tabIndex = 0;
signal View currentView = tabs.getViewByIndex(tabIndex);
```

The signal currentView is considered a signal of a signal because it contains an instance variable content, which is also a signal. Thus, the dependency between the inner signal content and currentView is constructed dynamically, i.e., only the content of the selected tab is redrawn (we will explain subscribe in the next section). The selection of the tab (i.e., the update of tabIndex) acts as the switch function of the traditional FRP.[6]

### 3.4 Events as Effects of Signals

SignalJ harmonizes the event mechanism with signals by considering that an event is a side effect of an update of a signal. Specifically, every signal in SignalJ can be an *effectful signal*. For example, we can implement an event handler that responds to an update in the signal. In SignalJ, an event handler is a lambda expression or method reference that is passed to the subscribe method called on the signal. The following code fragment shows an example:

```
signal int a = 5;
a.subscribe(e -> System.out.println(e));
a++; // display 6
```

The handler is called whenever the signal is updated. Thus, the lambda expression passed to subscribe is called at the subsequent a++, and the value of a, which is now 6, is displayed. The formal parameter represents the value of the signal when the handler is called.

This event system supports the fundamental characteristics of the existing event mechanisms. For example, we can register multiple event handlers by calling subscribe multiple times, which results in broadcasting when the corresponding signal is updated. SignalJ also supports event correlation and composition mechanisms as in EventJava [13], which are described in Section 3.6.

### 3.5 Pull Strategy for Signals, Push Strategy for Events

We have stated that both update propagation between signals and event handler invocation occur if the source signal is updated. However, we have not discussed when both are computed exactly. As we have discussed in Section 2.2, we may compute the signal update propagation either in a push- or pull-based manner, whereas the event handler is event-driven by definition.

In SignalJ, each signal is evaluated when it is accessed. This means that even though each composite signal comprises a static signal network, this demand-driven (pull) evaluation reflects the data dependency that is constructed dynamically (e.g., by calling a method embedded in the composite signal).

[6] We consider that signal fields can be a part of interface of the class, as a signal represents not only its state but its behavior. In this example, the user of class View should know that it contains the signal content.

On the other hand, each event handler is eagerly called when the subscribed signal is updated. The question is how to manage such a push-based invocation for a composite signal, which is evaluated on-demand. For this purpose, SignalJ maintains the signals on which the composite signal depends, and when one of these signals is updated, the event handler is called.

### 3.6 Event Composition

As in the existing event systems, a signal in SignalJ can be composed with other signals. The following code fragment shows examples:

```
signal boolean p = ..;
signal int s1 = ..;
signal int s2 = ..;
signal int s = s1.or(s2);
signal int ss = s1.when(p,0);
```

The signal s is a composite signal consisting of signals s1 and s2 that are composed using the or method, which returns a signal that is updated whether the method call receiver (s1) or argument (s2) is updated. The returned signal possesses the event sequence consisting of a merging of event sequences of the receiver and argument signals. Similarly, the when pseudo method returns a signal that possesses the event sequence of the receiver signal filtered by the argument predicate (p); that is, the update of ss occurs when s1 is updated and the predicate p is true. The second argument of when is a default value, which is necessary if p is false when ss is initialized.

This signal composition mechanism confers upon the signal mechanism the capability of composition provided by the event correlation mechanism [13, 33].

### 3.7 Implementation Notes

We implemented a proof-of-concept compiler of SignalJ. The compiler translates the SignalJ program into a Java program that uses the runtime library, and the runtime library implements the dynamic semantics such as update propagation and event subscription using RxJava2.

#### 3.7.1 Compiler

The SignalJ compiler is built as a module of ExtendJ, an extensible Java compiler implemented using JastAdd [10]. Basically, the compiler is implemented as a translation to Java 8 programs; then, the Java 8 compiler generates bytecode, which is executable on the standard Java virtual machine, from the translated programs. This generation of bytecode is automated by ExtendJ.

The translation proceeds as follows:

1. Identify all variables declared with signal

2. Perform type-checking regarding `signal`, e.g., check that imperative updates are performed only on source signals
3. Rewrite each `signal` declaration to a creation of a signal object, which is provided by the runtime library. Indeed, the annotation mechanism provided by SignalJ is implemented using lifting and unlifting; the right-hand side of `=` in each signal declaration is lifted to a signal. Internally, this signal is implemented using a `Flowable` (an RxJava2 class that provides the ability to consume reactive dataflows).
4. Rewrite each signal use to a method call on that signal that returns the current value of the signal (unlifting). Each imperative update of the source signal is also translated to a method call that sets a new value of the source signal.

#### 3.7.2 Runtime Library

The SignalJ runtime library is implemented as a wrapper for RxJava2. In brief, a source signal is implemented using `BehaviorProcessor`, which emits the most recent item it has observed. The method for updating the value of the source signal is implemented using the `onNext` method provided by `BehaviorProcessor`. Similarly, a composite signal is implemented using a `Flowable`, which is constructed by some operators such as `map` and `merge` (if the composite signal depends on multiple signals) called on the dependant signals. The event composition mechanism is also realized using such operators. Finally, the `subscribe` mechanism is naturally implemented using the subscription mechanism of RxJava2.

## 4 The Core Language

This section provides a formal model of SignalJ. In particular, this harmonization can be realized by understanding the following technical issues.

- Events and signals adopt different models of time progress and evaluation strategies. Thus, we need to specify the operational semantics that retains these features carefully.
- The annotation approach for representing signals simplifies the language mechanism because it makes the conversions between signals and non-signals unnecessary. However, there are no existing type system in this setting.

This section formalizes a core language of SignalJ as a small calculus, Featherweight SignalJ (FSJ), which is built on the basis of Featherweight Java (FJ) [19]. The ordering of signal propagation and event handlers, which is discussed in Section 2.2, is carefully encoded in the operational semantics of FSJ. We note that this calculus includes the side effects triggered by field assignment. Thus, unlike FJ, FSJ is not a functional calculus. In particular, FSJ focuses on the following fundamental features of SignalJ:

- Distinction between signals and other values in the type system
- Distinction between source and composite signals
- Interactions between push-based update propagation for calling event handlers and pull-based evaluation for signals
- Event handler registration using `subscribe`

**Table 1** Featherweight SignalJ: abstract syntax.

| | | | | | |
|---|---|---|---|---|---|
| CL | ::= | class C ◁ C { $\overline{\text{n C h=e}}$; $\overline{\text{n C f}}$; K; $\overline{\text{M}}$ } | e | ::= | x \| e.f \| e.m($\overline{\text{e}}$) \| new C($\overline{\text{e}}$) \| e.f = e \| e; e \| e.f.subscribe(e) \| $\ell$ \| $\{\text{e}\}_{\ell.\text{f}}$ \| $\epsilon$ \| let x=e in e |
| K | ::= | C($\overline{\text{C f}}$, $\overline{\text{C g}}$) { super($\overline{\text{g}}$); this.$\overline{\text{f}}$=$\overline{\text{f}}$; } | T | ::= | C \| Unit |
| M | ::= | T m($\overline{\text{C x}}$) { e } | | | |
| n | ::= | signal \| · | | | |

## 4.1 Syntax

Table 1 shows the abstract syntax of FSJ. Let the metavariables C, D, E, and F range over class names; n, o, and p range over modifiers; f, g, and h range over field names; m ranges over method names; d and e range over expressions; S, T, and U range over types; and x ranges over variables, which include a special variable, this. Overlines denote sequences, e.g., $\overline{\text{f}}$ stands for a possibly empty sequence $\text{f}_1, \cdots, \text{f}_n$. An empty sequence is denoted by •. We also use "this.$\overline{\text{f}}$=$\overline{\text{f}}$;" as shorthand for "this.$\text{f}_1$=$\text{f}_1$;$\cdots$;this.$\text{f}_n$=$\text{f}_n$;" where $n$ denotes the length of $\overline{\text{f}}$; "$\overline{\text{n C f}}$=$\overline{\text{e}}$;" as shorthand for "$\text{n}_1$ $\text{C}_1$ $\text{f}_1$=$\text{e}_1$;, $\cdots$ , $\text{n}_n$ $\text{C}_n$ $\text{f}_n$=$\text{e}_n$;"; and so on. We also use $\overline{\text{e}}$ as shorthand for $\text{e}_1$; $\cdots \text{e}_n$;. We use commas and semicolons for concatenations.

A class declaration consists of field declarations, a constructor declaration, and method declarations. A field declaration can be initialized with an expression. As shown in the type system, this field is supposed to be used to represent a composite signal, although this is not represented in the abstract syntax. A constructor initializes other fields, which include source signals, and those in the superclasses. An expression can be either a variable; a field access; a method invocation; a constructor invocation; a field assignment; a concatenation; a subscription; a location ($\ell$); a special expression, which can appear only in the reductions, of the form $\{\text{e}\}_{\ell.\text{f}}$, which represents a side effect under an assignment to the field $\ell$.f; an empty expression ($\epsilon$); or a let-bound expression. For simplicity, we do not assume a signal that is represented as a local variable. As FSJ represents an event in terms of a field assignment, which is used to produce a side effect and does not evaluate to a value, the empty expression is used to represent a special situation where there are no values. We also note that the argument to subscribe is not a lambda abstraction but is simplified as an expression. This expression represents a function that takes no parameters and returns no value. We formalize such "no value" as an empty expression $\epsilon$ that is never consumed. This simplifies the syntax of FSJ, and without loss of generality, we can formalize the mechanism of push-based calls of event handlers with this simplification.

There are two kinds of types in FSJ: class types and a unit type. A class type in FSJ is a normal class type that exists in Java and FJ. The unit type, Unit, is the type of the empty expression. It is also considered to represent the situation where there are no values (corresponding to "void" in Java), as the empty expression is never applied to other expressions.

A program ($CT$,e) consists of a class table $CT$ that maps a class name C to a class declaration CL and an expression e that corresponds to the body of the main method.

**Table 2** Featherweight SignalJ: auxiliary definitions

$$\boxed{\mathit{composite}(\mathsf{C}) = \overline{\mathsf{n}\ \mathsf{C}\ \mathsf{f=e}}}$$

$$\mathit{composite}(\mathsf{Object}) = \bullet$$

$$\frac{\mathsf{class\ C} \triangleleft \mathsf{D}\ \{\ \overline{\mathsf{n\ C\ f=e};}\ \cdots\ \} \qquad \mathit{composite}(\mathsf{D}) = \overline{\mathsf{o\ D\ g=d}}}{\mathit{composite}(\mathsf{C}) = \overline{\mathsf{o\ D\ g=d}}, \overline{\mathsf{n\ C\ f=e}}}$$

$$\boxed{\mathit{source}(\mathsf{C}) = \overline{\mathsf{n\ C\ f}}}$$

$$\mathit{source}(\mathsf{Object}) = \bullet$$

$$\frac{\mathsf{class\ C} \triangleleft \mathsf{D}\ \{\ \cdots\ \overline{\mathsf{n\ C\ f};}\ \cdots\ \} \qquad \mathit{source}(\mathsf{D}) = \overline{\mathsf{o\ D\ g}}}{\mathit{source}(\mathsf{C}) = \overline{\mathsf{o\ D\ g}}, \overline{\mathsf{n\ C\ f}}}$$

$$\boxed{\mathit{mbody}(\mathsf{m}, \mathsf{C}) = \overline{\mathsf{x}}.\mathsf{e}}$$

$$\frac{\mathsf{class\ C} \triangleleft \mathsf{D}\ \{\cdots\ \mathsf{C_0\ m}(\overline{\mathsf{C\ x}})\{\ \mathsf{e}\ \}\cdots\}}{\mathit{mbody}(\mathsf{m}, \mathsf{C}) = \overline{\mathsf{x}}.\mathsf{e}}$$

$$\frac{\mathsf{class\ C} \triangleleft \mathsf{D}\ \{\cdots\ \overline{\mathsf{M}}\ \} \qquad \mathsf{m} \notin \overline{\mathsf{M}}}{\mathit{mbody}(\mathsf{m}, \mathsf{C}) = \mathit{mbody}(\mathsf{m}, \mathsf{D})}$$

$$\boxed{\mathit{mtype}(\mathsf{m}, \mathsf{C}) = \overline{\mathsf{C}} \rightarrow \mathsf{T}}$$

$$\frac{\mathsf{class\ C} \triangleleft \mathsf{D}\ \{\cdots\ \mathsf{T\ m}(\overline{\mathsf{C\ x}})\{\ \mathsf{e}\ \}\cdots\}}{\mathit{mtype}(\mathsf{m}, \mathsf{C}) = \overline{\mathsf{C}} \rightarrow \mathsf{T}}$$

$$\frac{\mathsf{class\ C} \triangleleft \mathsf{D}\ \{\cdots\ \overline{\mathsf{M}}\ \} \qquad \mathsf{m} \notin \overline{\mathsf{M}}}{\mathit{mtype}(\mathsf{m}, \mathsf{C}) = \mathit{mtype}(\mathsf{m}, \mathsf{D})}$$

We assume that $CT(\mathsf{C})$ = class C ... for any $\mathsf{C} \in dom(CT)$, and no cycles exist in the transitive closure of $\triangleleft$ (extends). We also assume the conditions imposed by FJ that field hiding and method overloading are not allowed, and all fields in the same class and all parameters in the same method are distinct.

**Auxiliary definitions.** We provide some auxiliary definitions, which are shown in Table 2. There are two functions for looking up fields. The function $\mathit{composite}(\mathsf{C})$ returns a sequence $\overline{\mathsf{n\ C\ f=e}}$ of the quadruples of a field name, its type, its modifier, and its expression. This lookup function collects all of the composite signals from C and its superclasses (i.e., all initialized field declarations in C and its superclasses). Similarly, the function $\mathit{source}(\mathsf{C})$ returns a sequence $\overline{\mathsf{n\ C\ f}}$ of the triples of a field name, its type, and its modifier, which are initialized by the constructor. These fields include source signals and nonsignal fields.

The method body lookup function $\mathit{mbody}(\mathsf{m}, \mathsf{C})$ returns a pair $\overline{\mathsf{x}}.\mathsf{e}$ of parameters and an expression of method m in class C. The lookup proceeds to the superclass if m is not found in C. Similarly, the method type lookup function $\mathit{mtype}(\mathsf{m}, \mathsf{C})$ returns a pair $\overline{\mathsf{C}} \rightarrow \mathsf{T}$ of parameter types and a return type of method m in class C.

### 4.2 Operational Semantics

As we need to maintain the current value of each signal and its event handlers during the computation, it is necessary to provide runtime data structures for them. An object store $\mu$ is a mapping that maps a location $\ell$ to an object new $\mathsf{C}(\overline{\mathsf{v}})$. A handler store $\sigma$ is a mapping that maps a field access on a location $\ell.\mathsf{f}$ to an expression e, which represents an event handler registered to $\ell.\mathsf{f}$ by subscribe (as explained below,

**Table 3** Featherweight SignalJ: computation.

$$\frac{\mu(\ell) = \mathsf{new\ C}(\overline{\ell}) \qquad source(\mathsf{C}) = \overline{\mathsf{n\ C\ f}}}{\sigma \mid \mu \mid \ell.\mathsf{f}_i \longrightarrow \sigma \mid \mu \mid \ell_i} \quad \text{(R-FIELD)}$$

$$\frac{\mu(\ell) = \mathsf{new\ C}(\ldots) \qquad composite(\mathsf{C}) = \overline{\mathsf{n\ C\ f{=}e}}}{\sigma \mid \mu \mid \ell.\mathsf{f}_i \longrightarrow \sigma \mid \mu \mid \mathsf{e}_i} \quad \text{(R-FIELDS)}$$

$$\frac{\mu(\ell) = \mathsf{new\ C}(\overline{\ell'}) \qquad mbody(\mathsf{C},\mathsf{m}) = \overline{\mathsf{x}}.\mathsf{e}}{\sigma \mid \mu \mid \ell.\mathsf{m}(\overline{\ell}) \longrightarrow \sigma \mid \mu \mid [\overline{\ell}/\overline{\mathsf{x}}, \ell/\mathsf{this}]\mathsf{e}} \quad \text{(R-INVK)}$$

$$\sigma \mid \mu \mid \mathsf{new\ C}(\overline{\ell}) \longrightarrow \sigma \mid \mu \cup \{\ell \mapsto \mathsf{new\ C}(\overline{\ell})\} \mid \ell \quad \text{(R-NEW)}$$

$$\frac{\mu(\ell) = \mathsf{new\ C}(\overline{\ell}) \qquad source(\mathsf{C}) = \overline{\mathsf{n\ C\ f}} \qquad \mathsf{n}_i = \cdot \qquad \mu' = \mu \uplus \{\ell \mapsto [\ell'/\ell_i]\mathsf{new\ C}(\overline{\ell})\}}{\sigma \mid \mu \mid \ell.\mathsf{f}_i = \ell' \longrightarrow \sigma \mid \mu' \mid \epsilon} \quad \text{(R-ASSIGN)}$$

$$\frac{\mu(\ell) = \mathsf{new\ C}(\overline{\ell}) \qquad source(\mathsf{C}) = \overline{\mathsf{n\ C\ f}} \qquad \mathsf{n}_i = \mathsf{signal} \qquad \mu' = \mu \uplus \{\ell \mapsto [\ell'/\ell_i]\mathsf{new\ C}(\overline{\ell})\}}{\sigma \mid \mu \mid \ell.\mathsf{f}_i = \ell' \longrightarrow \sigma \mid \mu' \mid \{\sigma(\ell.\mathsf{f}_i)\}_{\ell.\mathsf{f}_i}} \quad \text{(R-ASSIGNS)}$$

$$\frac{handlers(\sigma, \mu, \ell.\mathsf{f}) = \overline{\mathsf{e}}}{\sigma \mid \mu \mid \{\epsilon\}_{\ell.\mathsf{f}} \longrightarrow \sigma \mid \mu \mid \overline{\mathsf{e}}} \quad \text{(R-ASSIGNCONT)}$$

$$\sigma \mid \mu \mid \ell.\mathsf{f.subscribe(e)} \longrightarrow \sigma \uplus \{\ell.\mathsf{f} \mapsto \sigma(\ell.\mathsf{f});\ \mathsf{e}\} \mid \mu \mid \epsilon \quad \text{(R-SUBSCRIBE)}$$

$$\sigma \mid \mu \mid \epsilon;\ \mathsf{e} \longrightarrow \sigma \mid \mu \mid \mathsf{e} \quad \text{(R-CAT)} \qquad \mathsf{let\ x{=}}\ell\ \mathsf{in\ e} \longrightarrow [\ell/\mathsf{x}]\mathsf{e} \quad \text{(R-LET)}$$

multiple event handlers registered to the same signal are concatenated using ; to form a single expression).

The operational semantics of FSJ, which is shown in Table 3, is given by a reduction relation of the form $\sigma \mid \mu \mid \mathsf{e} \longrightarrow \sigma' \mid \mu' \mid \mathsf{e}'$, which is read as "expression e under object store $\mu$ and handler store $\sigma$ reduces to $\mathsf{e}'$ under $\mu'$ and $\sigma'$." In the rules R-FIELD, R-FIELDS, R-ASSIGN, and R-ASSIGNS, we assume that $\overline{\mathsf{f}}$ and $\overline{\ell}$ have the same length $n$ and the index $i$ ranges over $1..n$. We also assume freshness conditions for some rules (for example, in R-NEW, $\ell$ is a fresh location that is not found in $\mu$).

**Pull-based evaluation of composite signals.** One of the notable computation rules is R-FIELDS, which defines the field access where the accessed field is a composite signal, as this ensures the immediate propagation of a change in the source signal. The composite signal is initialized with an expression $\mathsf{e}_i$, *which is evaluated every time the field is accessed*; that is, FSJ evaluates a composite signal using the pull strategy [11]. We note that the type system ensures that all $\mathsf{n}_i \in \overline{\mathsf{n}}$ in R-FIELDS are signal. R-FIELD defines the other case where the corresponding value of the field is taken from $\mu$. These

**Table 4** Featherweight SignalJ: side effect.

$$\boxed{\mathit{effect}(\mu, \ell.\mathsf{f}) = \overline{\ell.\mathsf{f}}}$$

$$\mathit{effect}(\emptyset, \ell.\mathsf{f}) = \bullet$$

$$\frac{\mathit{composite}(\mathsf{C}) = \overline{\mathsf{n}\ \mathsf{T}\ \mathsf{f=e}} \quad \mathit{effect}(\ell_0, \bar{\mathsf{f}}, \bar{\mathsf{e}}, \ell.\mathsf{f}) = \ell_0.\bar{\mathsf{g}} \quad \mathit{effect}(\mu, \ell_0.\bar{\mathsf{g}}) = \overline{\ell.\bar{\mathsf{f}}} \quad \mathit{effect}(\mu, \ell.\mathsf{f}) = \overline{\ell.\bar{\mathsf{f}}}'}{\mathit{effect}(\{\ell_0 \mapsto \mathsf{new}\ \mathsf{C}(\bar{\ell})\} \cup \mu, \ell.\mathsf{f}) = \overline{\ell.\bar{\mathsf{f}}}, \ell_0.\bar{\mathsf{g}}, \overline{\ell.\bar{\mathsf{f}}}'}$$

$$\boxed{\mathit{effect}(\ell_0, \bar{\mathsf{f}}, \bar{\mathsf{e}}, \ell.\mathsf{f}) = \ell_0.\bar{\mathsf{f}}}$$

$$\mathit{effect}(\ell_0, \bullet, \bullet, \ell.\mathsf{f}) = \bullet$$

$$\frac{\mathit{contains}([\ell_0/\mathsf{this}]\mathsf{e}, \ell.\mathsf{f})}{\mathit{effect}(\ell_0, \mathsf{f}_i\bar{\mathsf{f}}, \mathsf{e}\bar{\mathsf{e}}, \ell.\mathsf{f}) = \ell_0.\mathsf{f}_i, \mathit{effect}(\ell_0, \bar{\mathsf{f}}, \bar{\mathsf{e}}, \ell.\mathsf{f})}$$

$$\frac{\mathit{contains}([\ell_0/\mathsf{this}]\mathsf{e}, \ell.\mathsf{f}) \text{ undefined}}{\mathit{effect}(\ell_0, \mathsf{f}_i\bar{\mathsf{f}}, \mathsf{e}\bar{\mathsf{e}}, \ell.\mathsf{f}) = \mathit{effect}(\ell_0, \bar{\mathsf{f}}, \bar{\mathsf{e}}, \ell.\mathsf{f})}$$

$$\boxed{\mathit{handlers}(\sigma, \mu, \ell.\mathsf{f}) = \bar{\mathsf{e}}}$$

$$\frac{\mathit{effect}(\mu, \ell.\mathsf{f}) = \overline{\ell.\bar{\mathsf{f}}} \qquad \mathsf{e}_k = \sigma(\ell_i.\mathsf{f}_{ij})}{\mathit{handlers}(\sigma, \mu, \ell.\mathsf{f}) = \bar{\mathsf{e}}}$$

$$\boxed{\mathit{contains}(\mathsf{e}, \ell.\mathsf{f})}$$

$$\mathit{contains}(\ell.\mathsf{f}, \ell.\mathsf{f})$$

$$\frac{\mathit{contains}(\mathsf{e}, \ell.\mathsf{f})}{\mathit{contains}(\mathsf{e.g}, \ell.\mathsf{f})}$$

$$\frac{\mathit{contains}(\mathsf{e}_0, \ell.\mathsf{f}) \vee \cdots \vee \mathit{contains}(\mathsf{e}_n, \ell.\mathsf{f})}{\mathit{contains}(\mathsf{e}_0.\mathsf{m}(\bar{\mathsf{e}}), \ell.\mathsf{f})}$$

$$\frac{\mathit{contains}(\mathsf{e}_1, \ell.\mathsf{f}) \vee \cdots \vee \mathit{contains}(\mathsf{e}_n, \ell.\mathsf{f})}{\mathit{contains}(\mathsf{new}\ \mathsf{C}(\bar{\mathsf{e}}), \ell.\mathsf{f})}$$

field values can be changed by field assignment, which immediately changes $\mu$ using the assigned value. The field assignment is divided into two rules: R-Assign and R-AssignS. R-Assign is the case where the field is not a signal, and the assignment expression simply reduces to an empty expression. R-AssignS is the case where the field is a (source) signal and the event handler calls follow this reduction (see below). In R-Assign, R-AssignS and R-Subscribe, we use $\uplus$ as a destructive update of the set; that is, $(x \uplus y)(k) = y(k)$ if $k \in \mathit{dom}(y)$ or $x(k)$ otherwise. $[\ell'/\ell_i]\mathsf{new}\ \mathsf{C}(\bar{\ell})$ denotes that the $i$th argument of the instance of C, i.e., the value of the $i$th field of that instance, is replaced with $\ell'$. These rules ensure that each source signal is updated when the field assignment is performed, and the value of a composite signal is always computed using these up-to-date source signal values.

**Push-based invocation of event handlers.** The evaluation of event handlers is performed in the push strategy [11]; that is, event handlers must be invoked whenever an update of the signal is carried on. In particular, when a source signal is updated, all event handlers registered to that signal *and all dependent composite signals* are evaluated. This invocation of event handlers is two-fold.

First, the event handlers $\sigma(\ell.\mathsf{f}_i)$ registered to the source signal are invoked (R-AssignS). As R-Subscribe specifies, subscribe concatenates the argument e at the tail of the existing handlers $\sigma(\ell.\mathsf{f})$. Thus, $\sigma(\ell.\mathsf{f})$ always returns a single expression (if $\sigma(\ell.\mathsf{f})$ is not defined, it returns an empty expression). The braces { e } obtained by R-AssignS are reduced to $\{\epsilon\}$ (if e is not $\epsilon$) by the congruence rule shown below, which is further evaluated by R-AssignCont that executes event handlers obtained by *handlers* (as mentioned above, we use $\overline{\mathsf{e}}$ as shorthand for $\mathsf{e}_1;\cdots;\mathsf{e}_n$). We note that the type system ensures that the modifier of f in R-Subscribe is signal.

Then, the event handlers registered to all dependent composite signals of that source signal, which are obtained by function *handlers*$(\sigma,\mu,\ell.\mathsf{f})$, are invoked. *handlers*$(\sigma,\mu,\ell.\mathsf{f})$, which is read as "event handlers of all sinks of $\ell.\mathsf{f}$ that are registered to $\sigma$," is defined in Table 4. This function returns an expression e constructed by concatenating all of the event handlers of "all sinks of $\ell.\mathsf{f}$ in $\mu$," which is expressed as *effect*$(\mu,\ell.\mathsf{f})$. The function *effect*$(\mu,\ell.\mathsf{f})$ recursively searches composite signals that contain $\ell.\mathsf{f}$ and its sinks. The overloaded function *effect*$(\ell_0,\overline{\mathsf{f}},\overline{\mathsf{e}},\ell.\mathsf{f})$ determines which expression in $\overline{\mathsf{e}}$ contains $\ell.\mathsf{f}$ (the first and second arguments $\ell_0$ and $\overline{\mathsf{f}}$ are used to construct a result). The function *contains*$(\mathsf{e},\ell.\mathsf{f})$ specifies the lexical-scope-based event handler lookup; it is defined if $\ell.\mathsf{f}$ is contained in any of the subexpressions of e. It recursively searches $\ell.\mathsf{f}$ for all of e's subexpressions. The type system ensures that *contains* is applied to expressions consisting of only field accesses, method invocations, and instance creations (T-Class). Thus, in R-AssignCont, all event handlers registered to the composite signals that depend on $\ell.\mathsf{f}$ are evaluated.

We note that *effect* does not determine the order of event handlers. The implementors may determine this ordering according to their own convenience (e.g., event handlers may be invoked in parallel).

**Congruence rules.** To show the congruence rule that enables a reduction of subexpressions, we introduce the evaluation context $E$ as follows:

$$E \ ::= \ [\,].\mathsf{f} \mid [\,].\mathsf{m}(\overline{\mathsf{e}}) \mid \ell.\mathsf{m}(\overline{\ell},[\,],\overline{\mathsf{e}}) \mid \mathsf{new}\ \mathsf{C}(\overline{\ell},[\,],\overline{\mathsf{e}}) \mid [\,].\mathsf{f} = \mathsf{e} \mid \ell.\mathsf{f} = [\,] \mid \{[\,]\}_{\ell.\mathsf{f}} \mid [\,];\, \mathsf{e} \mid [\,].\mathsf{f.subscribe}(\mathsf{e}) \mid \mathsf{let}\ \mathsf{x}=[\,]\ \mathsf{in}\ \mathsf{e}$$

Each evaluation context is an expression with a hole (written []) somewhere inside it. We write $E[\mathsf{v}]$ for an expression obtained by replacing the hole in $E$ with v.

The congruence rule is defined as follows:

$$\frac{\sigma \mid \mu \mid \mathsf{e} \longrightarrow \sigma \mid \mu \mid \mathsf{e}'}{\sigma \mid \mu \mid E[\mathsf{e}] \longrightarrow \sigma \mid \mu \mid E[\mathsf{e}']}$$

The evaluation context syntactically ensures that the reduction of the right-hand side of ; occurs after the left-hand side reduces to an empty value (R-Cat), and the argument of subscribe is not reduced when it is called. It also defines the evaluation order of arguments to method and constructor invocations.

### 4.3 Type System

The computation rules do not use any special forms to represent signals at runtime. Thus, the calculus does not provide any operations to convert signals to non-signals and vice versa. This simplifies the calculus, as we need not provide computation rules for the lifting and unlifting of signals. In this sense, FSJ formalizes signals and events in a different way from the traditional FRP approach [12].

In this setting, we show that FSJ is also type-safe. Intuitively, the annotation signal is used to check that the operation subscribe is applied only to signals and to statically construct signal networks consisting of source and composite signals, which are object fields.

Despite this difference, the type system of FSJ is a straightforward adaptation of that of FJ [19], except that we need to handle the typing of runtime values, i.e., locations, which appear in the reduction rules. We also need to provide additional typing rules for expressions that are introduced by FSJ such as $\{\mathtt{e}\}_{\ell.\mathtt{f}}$ and subscribe.

The subtyping relation $\mathtt{S} <: \mathtt{T}$ is defined as the reflexive and transitive closure of the extends ($\triangleleft$) clauses:

$$\mathtt{T} <: \mathtt{T} \qquad \frac{\mathtt{U} <: \mathtt{S} \quad \mathtt{S} <: \mathtt{T}}{\mathtt{U} <: \mathtt{T}} \qquad \frac{\mathtt{class}\ \mathtt{C} \triangleleft \mathtt{D}\,\{\cdots\}}{\mathtt{C} <: \mathtt{D}}$$

We note that there are no subtypes of Unit, and Unit cannot be a subtype of any types other than itself, as a value of type Unit, $\epsilon$, is not supposed to be an argument to method and constructor invocation, and the left-hand side of an assignment (=).

The typing rules for expressions are shown in Table 5. A type environment $\Gamma$ is a finite mapping from variables to class names. A store typing $\Sigma$ is a finite mapping from locations to their types. A type judgment for expressions is of the form $\Gamma \mid \Sigma \vdash \mathtt{e} : \mathtt{C}$, read as "expression e is given type C under type environment $\Gamma$ and store typing $\Sigma$."

The first four rules, T-VAR for variables, T-FIELD for field access, T-INVK for method invocation, and T-NEW for constructor invocation, are straightforward adaptations of those of FJ. Note that a variable x can only be given a class type. A field assignment expression is given the type Unit if the receiver is given a type, field $\mathtt{f}_i$ is found in that type, and the type of the assigned expression is a subtype of this field. In addition, this assignment is allowed only if $\mathtt{f}_i$ is a source signal or nonsignal, which is ensured by using *source* for field lookup. A runtime expression $\{\mathtt{e}\}_{\ell.\mathtt{f}}$ is given the type Unit only when the enclosed expression e is given the type Unit. A concatenate expression is given a type only when the type of the left-hand side of ; is Unit. A subscribe expression is also given the type Unit if the field f is found in the receiver type, the declared type of f is a signal, and e is given the type Unit. The auxiliary function *ftype*(C, f) returns a type of f using *composite* and *source*, which is defined as follows:

$$\frac{\mathit{composite}(\mathtt{C}) = \overline{\mathtt{n}\ \mathtt{C}\ \mathtt{f}\texttt{=}\mathtt{e}}}{\mathit{ftype}(\mathtt{C},\mathtt{f}_i) = \mathtt{n}_i\ \mathtt{C}_i} \qquad \frac{\mathit{source}(\mathtt{C}) = \overline{\mathtt{n}\ \mathtt{C}\ \mathtt{f}}}{\mathit{ftype}(\mathtt{C},\mathtt{f}_i) = \mathtt{n}_i\ \mathtt{C}_i}$$

The method and class typing are shown in Table 6. A type judgment for methods in a class is of the form M ok in C, read as "method M is well-formed in class C." This judgment is represented by the typing rule T-METHOD, which checks that the method body is given a subtype of the declared type. A class is well-formed (written CL ok) if the constructor matches the field declarations other than composite signals, all composite signals have a signal type, all expressions $\overline{\mathtt{e}}$ initializing the composite signals

**Table 5** Featherweight SignalJ: expression typing

$$\frac{\Gamma = \overline{x} : \overline{C}}{\Gamma \mid \Sigma \vdash x_i : C_i} \;(\text{T-Var}) \qquad \frac{\Gamma \mid \Sigma \vdash e : \mathsf{Unit}}{\Gamma \mid \Sigma \vdash \{e\}_{\ell.f} : \mathsf{Unit}} \;(\text{T-AssignCont})$$

$$\frac{\Gamma \mid \Sigma \vdash e_0 : C_0 \qquad \mathit{ftype}(C_0, f) = n\ C}{\Gamma \mid \Sigma \vdash e_0.f : C} \;(\text{T-Field})$$

$$\frac{\Gamma \mid \Sigma \vdash e_0 : C_0 \qquad \Gamma \mid \Sigma \vdash \overline{e} : \overline{E} \qquad \overline{E} <: \overline{D} \qquad \mathit{mtype}(m, C_0) = \overline{D} \to T}{\Gamma \mid \Sigma \vdash e_0.m(\overline{e}) : T} \;(\text{T-Invk})$$

$$\frac{\mathit{source}(C_0) = \overline{n\ C\ f} \qquad \Gamma \mid \Sigma \vdash \overline{e} : \overline{D} \qquad \overline{D} <: \overline{C}}{\Gamma \mid \Sigma \vdash \mathsf{new}\ C_0(\overline{e}) : C_0} \;(\text{T-New})$$

$$\frac{\Gamma \mid \Sigma \vdash e_0 : C_0 \qquad \mathit{source}(C_0) = \overline{n\ C\ f} \qquad \Gamma \mid \Sigma \vdash e : D \qquad D <: C_i}{\Gamma \mid \Sigma \vdash e_0.f_i = e : \mathsf{Unit}} \;(\text{T-Assign})$$

$$\frac{\Gamma \mid \Sigma \vdash e : \mathsf{Unit} \qquad \Gamma \mid \Sigma \vdash e' : T}{\Gamma \mid \Sigma \vdash e; e' : T} \;(\text{T-Cat})$$

$$\frac{\Gamma \mid \Sigma \vdash e_0 : C_0 \qquad \mathit{ftype}(C_0, f) = \mathsf{signal}\ C \qquad \Gamma \mid \Sigma \vdash e : \mathsf{Unit}}{\Gamma \mid \Sigma \vdash e_0.f.\mathsf{subscribe}(e) : \mathsf{Unit}} \;(\text{T-Subscribe})$$

$$\frac{\Gamma \mid \Sigma \vdash e : C \qquad \Gamma, x : D \mid \Sigma \vdash e_0 : T \qquad C <: D}{\Gamma \mid \Sigma \vdash \mathsf{let}\ x\text{=}e\ \mathsf{in}\ e_0 : T} \;(\text{T-Let})$$

$$\frac{\Sigma(\ell) = C}{\Gamma \mid \Sigma \vdash \ell : C} \;(\text{T-Loc}) \qquad \frac{}{\Gamma \mid \Sigma \vdash \epsilon : \mathsf{Unit}} \;(\text{T-Empty})$$

**Table 6** Featherweight SignalJ: method and class typing

$$\frac{\overline{x} : \overline{C}, \mathsf{this} : C \mid \emptyset \vdash e_0 : S \qquad S <: T}{T\ m(\overline{C\ x})\ \{\ e_0\ \}\ \text{ok in}\ C} \;(\text{T-Method})$$

$$\frac{\begin{array}{cccc} K = C(\overline{C\ f}, \overline{E\ g})\{\ \mathsf{super}(\overline{g}); \mathsf{this}.\overline{f}\text{=}\overline{f};\ \} & \mathsf{this} : C \mid \emptyset \vdash \overline{e} : \overline{F} & \overline{F} <: \overline{D} & \mathit{init}(\overline{e}) \\ \forall \overline{n}.n_i = \mathsf{signal} \quad D\ \text{ok} & \mathit{source}(D) = \overline{p\ E\ g} & \overline{M}\ \text{ok in}\ C & \end{array}}{\mathsf{class}\ C \lhd D\ \{\ \overline{n\ D\ h\text{=}e};\ \overline{o\ C\ f};\ K\ \overline{M}\ \}\ \text{ok}} \;(\text{T-Class})$$

are well-formed field initializers, the superclass is well-formed, and all methods are well-formed. We note that the well-formedness of the field initializers is defined as follows, which ensures that *contains* in Table 4 is defined over field initializers:

$$\mathit{init}(\mathsf{x}) \qquad \frac{\mathit{init}(\mathsf{e})}{\mathit{init}(\mathsf{e.f})} \qquad \frac{\mathit{init}(\mathsf{e}_0) \vee \cdots \vee \mathit{init}(\mathsf{e}_n)}{\mathit{init}(\mathsf{e}_0.\mathsf{m}(\overline{\mathsf{e}}))} \qquad \frac{\mathit{init}(\mathsf{e}_1) \vee \cdots \vee \mathit{init}(\mathsf{e}_n)}{\mathit{init}(\mathsf{new}\ \mathsf{C}(\overline{\mathsf{e}}))}$$

**Properties of the type system.** We show that FSJ is type safe as expected. To state the type safety, we first need to provide additional definitions regarding the well-typedness of an object store and a handler store. Proofs are shown in Appendix A.

**Definition 1.** *A store $\mu$ is said to be* well-typed *with respect to a typing context $\Gamma$ and a store typing $\Sigma$, written $\Gamma \mid \Sigma \vdash \mu$, if $dom(\mu) = dom(\Sigma)$ and $\Gamma \mid \Sigma \vdash \mu(\ell) : \Sigma(\ell)$ for every $\ell \in dom(\mu)$.*

**Definition 2.** *A handler store $\sigma$ is said to be* well-typed *with respect to a typing context $\Gamma$ and a store typing $\Sigma$, written $\Gamma \mid \Sigma \vdash \sigma$, if $\forall e \in codom(\sigma).\Gamma \mid \Sigma \vdash e : \mathit{Unit}$ and $\forall \ell.f \in dom(\sigma).\ell \in dom(\Sigma)$.*

The type safety of FSJ is obtained by the following theorems:

**Theorem 1** (Subject Reduction)**.** *Suppose that $\forall CL \in dom(CT).CL$ ok. If $\Gamma \mid \Sigma \vdash e : T$, $\Gamma \mid \Sigma \vdash \mu$, $\Gamma \mid \Sigma \vdash \sigma$, and $\sigma \mid \mu \mid e \longrightarrow \sigma' \mid \mu' \mid e'$, then $\Gamma \mid \Sigma' \vdash e' : S$, $\Gamma \mid \Sigma' \vdash \sigma'$ and $\Gamma \mid \Sigma' \vdash \mu'$ for some $\Sigma' \supseteq \Sigma$ and $S$ such that $S <: T$.*

**Theorem 2** (Progress)**.** *Suppose that $\emptyset \mid \Sigma \vdash e : T$ for some $T$ and $\Sigma$. Then, either $e$ is a location or an empty expression, or, for any store $\mu$ such that $\emptyset \mid \Sigma \vdash \mu$, there are some expression $e'$ and store $\mu'$ such that $\sigma \mid \mu \mid e \longrightarrow \sigma' \mid \mu' \mid e'$ and $\Gamma \mid \Sigma' \vdash \mu'$ for some $\Sigma' \supseteq \Sigma$, $\sigma$, and $\sigma'$ where $\sigma' \supseteq \sigma$.*

This type safety can be a basis for proving other properties regarding the behavior of signals and events. In particular, our type system does not distinguish signals from non-signals when their values are accessed. Nevertheless, our system still knows whether the field access is a signal (as shown by T-Field). We show some basic properties to ensure that the program behaves as expected in this setting. First, the following theorem ensures that an assignment to a non-signal does not trigger any event handler invocation (the proof is immediate from R-Assign).

**Theorem 3.** *If $\sigma \mid \mu \mid \ell.f = \ell' \longrightarrow \sigma' \mid \mu' \mid e$, $\Gamma \mid \Sigma \vdash \ell : C_0$, and $ftype(C_0, f) = \cdot\ C$, then $e = \epsilon$.*

The following theorem ensures that a subscriber $\mathsf{e}$ of field $\mathsf{f}$ is called when its source $\ell.\mathsf{g}$ is updated (the proof is shown in Appendix A).

**Theorem 4.** *If $e$ contains $\ell_0.f.subscribe(e_0)$ as a subexpression, $\sigma \mid \mu \mid e \longrightarrow^* \sigma' \mid \mu' \mid \{\sigma'(\ell.g)\}_{\ell.g}$, and $\ell_0.f \in effect(\mu', \ell.g)$, then $e \in handlers(\sigma', \mu', \ell.g)$.*

We note that this theorem states nothing regarding whether $\mathsf{f}$ is a signal. The type system ensures that $\mathsf{f}$ is a signal if `subscribe` is called on that.

We also state that reevaluation of a composite signal is ensured (the proof is immediate from R-FieldS).

**Theorem 5.** *If* $\Gamma \mid \Sigma \vdash \ell : C_0$*, signal* $C\, f = e \in composite(C_0)$*, and* $\sigma \mid \mu \mid \ell.f \longrightarrow \sigma' \mid \mu' \mid e'$*, then* $\sigma = \sigma'$*,* $\mu = \mu'$*, and* $e = e'$*.*

Finally, we show that in a well-typed program, reassignment for a composite signal never occurs during the computation; that is, if a well-typed expression is reduced to a field assignment, this field is not a composite signal.

**Theorem 6.** *If* $\Gamma \vdash \Sigma \vdash e : T$*,* $\sigma \mid \mu \mid e \longrightarrow \sigma' \mid \mu' \mid \ell.f = \ell'$*,* $\Gamma \mid \Sigma' \vdash \ell : C_0$ *for some* $\Sigma' \supseteq \Sigma$*, and* $composite(C_0) = \overline{n\ C\ f}$*, then* $f \notin \bar{f}$*.*

*Proof.* By Theorem 1 and T-Assign, $f \in \bar{g}$ where $source(C_0) = \overline{o\ D\ g}$. As $\bar{f} \cap \bar{g} = \bullet$, we have $f \notin \bar{f}$. □

**Prospection.** As these theorems only describe the basic properties about events and signals, more advanced features regarding event-signal interactions, such as the absence of infinite loops, are not discussed. For example, the type system does not detect a situation where a signal triggers an event that assigns to that signal again. We consider that it is possible to develop a loop detector based on the calculus by applying existing static analysis mechanisms such as an interprocedural control-flow analysis and a call-graph construction with a class-hierarchy analysis to the event handler. Even though such an analysis would require the entire program if naively applied, we may avoid an analysis of the external code such as the libraries and frameworks with the assumption that the external code will not directly use the user-defined signals [2]. As the signals on which the event handler depends are statically determined (by *effect* in Table 4), we prospect that this construction of a conservative analysis would be relatively easy.

## 5 Related Work

In addition to the closely related work discussed in Section 2, there are several pieces of work related to our approach that are worth discussing in this paper.

**Synchronous and FRP languages.** The techniques for reactive programming have been inspired by two directions of work: synchronous and FRP languages. Synchronous languages such as Lustre [17], Esterel [4], and Lucid Synchrone [31] view programs as fixed networks of stream processing nodes that communicate with each other. To represent a data flow on the basis of a synchronous notion of time, these languages define nets of operators connected with wires. These languages mainly focus on real-time requirements such as bounded memory and time consumption.

The study of FRP originated with Elliott and Hudak [12], who focused on time-varying values in functional programs. FRP provides a model that is richer than that of synchronous languages; that is, dataflow networks can dynamically grow and change. One use of FRP is similar to that of data-flow languages: to combine host language components for some methods that provide well-defined resource use characteristics. RT-FRP [37] is a statically typed language, which deals with the use of FRP in real-time applications and identifies a subset of FRP where it can statically guarantee that the time and space costs for a given program are statically bounded. E-FRP [38] further simplifies this idea by generalizing the global clock used in RT-FRP to a set of events. On the other hand, it has been proven that FRP-like dynamic reactive applications can be constructed on the basis of the efficient stateful execution model that synchronous languages provide [22]. A higher-order FRP language that supports an efficient implementation strategy to avoid spacetime leaks by the eager deallocation of old values was also proposed [21]. The push-pull FRP has also been proposed [11]. The purpose is slightly different from our approach. In the push-pull FRP, the evaluation strategy is determined for the efficient implementation to combine both benefits. On the other hand, our proposal combines the push and pull strategies to combine imperative events with signals.

FRP features are also embedded in general-purpose languages. This direction includes AFRP [9] and Yampa [29]. Frappé [8] is a Java library for realizing FRP, whose idea was originally implemented in Haskell. ReactiveML [26] is a reactive programming language based on ML that combines the synchronous model of Esterel with asynchronous models such as the dynamic creation of processes based on Boussinot's reactive model [5]. Ignatoff et al. proposed a method to adapt existing libraries written in OO languages (such as GUI libraries) to the FRP language FrTime [20]. Seamless integration is supported by several useful abstractions based on mixins, and these patterns are represented using macros to avoid code duplication.

When we focus on the propagation between signals in SignalJ (besides the event-based features), SignalJ appears to be a synchronous language at first, as each object field appears to be a static network. However, by adopting the OO feature of object creation, SignalJ can dynamically compose such a network (we note that each field is evaluated on-demand). A notable feature is its simplified approach where signals and non-signals are uniformly treated; that is, the construct `signal` is not a type constructor but a modifier. To the best of our knowledge, this simplification is not found in other existing OO-based FRP languages such as Microsoft's Reactive Extensions (Rx), Scala.React [23], REScala [33], DREAM [27], and the React JavaScript library.

**Event-based languages.** The next related research field is about event-based languages. As mentioned in Section 2, there are several event-based languages. EventJava [13] is an extension of Java with event-based mechanisms such as event matching, predicate guarding, event compositions and correlation. Ptolemy [32] is another event-based language that supports event subscriptions. More complex mechanisms such as the composition of event sequences [25, 24] have also been proposed.

Unlike these languages, SignalJ does not provide events as first-class constructs. Instead, in SignalJ, events are completely unified with signals. A notable finding is that we can still provide the fundamental features of events such as event handlers and event correlations in terms of signals.

The handling of side effects in FRP is also related to event-based language mechanisms (that are unified with signals) in that event handlers are usually used to produce side effects. This handling of side effects in FRP was initiated by Patai [30], where Elerea, a discrete time FRP implementation without the notion of event-based switch-ing, was proposed. Elerea provides stateful signals, where how the signal changes its value is specified. We can represent signals with side effects through a monadic interface where the value of a signal is accompanied by effectful computation. Instead of this functional setting, SignalJ is based on an imperative setting where the side effects originate from assignments to source signals.

**Other computational models.** Schuster proposed reactive variables [34] as a reactive extension of imperative languages. In his work, he introduced the rlet construct, which is similar to signal in SignalJ, to JavaScript. A lambda calculus based operational semantics is also illustrated, though proofs of some important properties including the type safety are not presented.

Constraint programming is similar to SignalJ in that both consider automatic propagation of changes, though in constraint programming, this propagation is taken bidirectionally. It supports declarative relations between program entities and automatic enforcement of their consistency. For example, Kaleidoscope [15] gives the user an introduction to constraints, which are automatically satisfied by the framework. Babersberg/JS [14] is an object constraint language that is based on JavaScript. In such a language, constraint solvers are used to update dependencies; SignalJ-like abstractions for time-varying computations are not provided.

Finally, some approaches to incremental computing (IC) that utilize data dependency are closely related to FRP and thus worth discussing with regards to their relation to SignalJ. IC uses memoization to avoid unnecessary recomputation when an input changes. Self-adjusting computation [1] is an approach to IC that is applied to a functional programming language and uses a special form of memoization where some portions of dynamic dependence graphs are cached and reused. This is implemented as an ML library. Carlsson translated this library into a Haskell library where the type system ensures the correct usage of that library [6]. ADAPTON [18] also uses such dependencies but the propagations are computed in an on-demand manner to provide efficient support of recomputation.

SignalJ is not concerned with the incremental changes studied in IC in that programmers do not have to program with incremental changes explicitly. For example, in SignalJ, it is not necessary for programmers to control which portions of the program should be evaluated eagerly. We have shown that the simple annotation approach to declare a signal is applied in this setting. Even though there may be a performance problem (for example, the SignalJ reevaluation rule requires that the entire expression is reevaluated when the composite signal is accessed), our language design significantly simplifies the program model. We consider that some portions of IC-based techniques might be applied to a more efficient implementation of SignalJ.

# A Proofs

We first show the lemmas required in the proof of Theorem 1.

**Lemma 1** (Substitution). *If* $\Gamma, \overline{x} : \overline{C} \mid \Sigma \vdash e_0 : T$ *and* $\Gamma \mid \Sigma \vdash \overline{\ell} : \overline{D}$ *and* $\overline{D}$ <: $\overline{C}$, *then* $\Gamma \mid \Sigma \vdash [\overline{\ell}/\overline{x}]e_0 : S$ *and* $S$ <: $T$ *for some* $S$.

*Proof.* By induction on $\Gamma, \overline{x} : \overline{C} \vdash e_0 : T$. □

**Lemma 2.** *Suppose* $mbody(m, C) = \overline{x}.e_0$ *and* $mtype(m, C) = \overline{D} \to T$. *Then,* $\overline{x} : \overline{D}, this : C' \vdash e_0 : S$ *for some* $C'$ *and* $S$ *such that* $C$ <: $C'$ *and* $S$ <: $T$.

*Proof.* By induction on $mbody(m, C) = \overline{x}.e_0$. □

We show the proof of Theorem 1 as follows.

*Proof.* By induction on $\sigma \mid \mu \mid e \longrightarrow \sigma' \mid \mu' \mid e'$ with the case analysis on the last reduction rule used.

**Case** R-Field:

$e = \ell.f_i \qquad e' = \ell_i$

By T-Field, $\Gamma \mid \Sigma \vdash \ell : C_0$ and $ftype(C_0, f_i) = n\ C$ for some n and $C_0$. As $\Gamma \mid \Sigma \vdash \mu$, $\Gamma \mid \Sigma \vdash \mu(\ell) : \Sigma(\ell)$. By T-New and T-Loc, we have $source(C_0) = \overline{n\ C\ f}$, $\Gamma \mid \Sigma \vdash \overline{e} : \overline{D}$, and $\overline{D}$ <: $\overline{C}$. Letting $D_i = D$ finishes the case.

**Case** R-FieldS:

$e = \ell.f_i \qquad e' = e_i$

By T-Field, $\Gamma \mid \Sigma \vdash \ell : C_0$ and $ftype(C_0, f_i) = n\ C$ for some n and $C_0$. As $\Gamma \mid \Sigma \vdash \mu$, $\Gamma \mid \Sigma \vdash \mu(\ell) : \Sigma(\ell)$. By the fact that $\forall CL \in dom(CT).CL$ ok, we have $composite(C_0) = \overline{n\ C\ f=e}$, $\Gamma \mid \Sigma \vdash \overline{e} : \overline{D}$, and $\overline{D}$ <: $\overline{C}$. Letting $D_i = D$ finishes the case.

**Case** R-Invk:

$e = \ell.m(\overline{\ell}) \qquad e' = [\overline{\ell}/\overline{e}, \ell/this]e_0 \qquad mbody(C_0, m) = \overline{x}.e_0 \qquad \mu(\ell) = new\ C_0(\overline{\ell}')$

By T-Invk, $\Gamma \mid \Sigma \vdash \ell.m(\overline{\ell}) : T$, $\Gamma \mid \Sigma \vdash \ell : C_0$, $\Gamma \mid \Sigma \vdash \overline{\ell} : \overline{E}$, $\overline{E}$ <: $\overline{D}$, $mtype(m, C_0) = \overline{D} \to T$. By Lemma 2, $\Gamma \mid \Sigma \vdash e_0 : U$ for some U such that U <: T. By Lemma 1, $\Gamma \mid \Sigma \vdash [\overline{\ell}/\overline{x}]e_o : S$ for some S such that S <: U. By the transitivity of the subtyping rule, we have S <: T, finishing the case.

**Case** R-New:

$e = new\ C(\overline{\ell}) \qquad e' = \ell \qquad \mu' = \mu \cup \{\ell \mapsto new\ C(\overline{\ell})\}$

By T-New, $\Gamma \mid \Sigma \vdash new\ C(\overline{\ell}) : C$. Letting $\Sigma' = \Sigma \cup \{\ell \mapsto C\}$ finishes the case.

**Case** R-Assign

$e = \ell.f_i = \ell' \qquad e' = \epsilon \qquad \mu' = \mu \cup \{\ell \mapsto [\ell'/\ell_i]new\ C(\overline{\ell})\}$

By T-Assign, $\Gamma \mid \Sigma \vdash \ell.f_i = \ell' : Unit$. It is easy to show that $\Gamma \mid \Sigma \vdash \mu'$. Then, T-Empty finishes the case.

**Case** R-AssignS:

$e = \ell.f_i = \ell' \qquad e' = \{\sigma(\ell.f_i)\}_{\ell.f_i} \qquad \mu' = \mu \cup \{\ell \mapsto [\ell'/\ell_i]new\ C(\overline{\ell})\}$

By T-Assign, $\Gamma \mid \Sigma \vdash \ell.f_i = \ell' : Unit$. It is easy to show that $\Gamma \mid \Sigma \vdash \mu'$. By the well-formedness of $\sigma$, $\Gamma \mid \Sigma \vdash \sigma(\ell.f_i) : Unit$. Then, T-AssignCont finishes the case.

**Case** R-AssignCont:

$e = \{\epsilon\}_{\ell.f} \qquad e' = e_0 \qquad handlers(\sigma, \mu, \ell.f) = e_0$

By T-AssignCont, $\Gamma \mid \Sigma \vdash \{\epsilon\}$ : Unit. By the fact that $\forall$CL $\in$ *dom*(*CT*).CL ok and the definition of *handlers*, it is easy to show that T-Cat finishes the case.

**Case** R-Cat: Immediate from T-Cat.

**Case** R-Publish:

$$\mathsf{e} = \ell.\mathsf{f}.\mathsf{publish}(\mathsf{e}_0) \qquad \mathsf{e}' = \epsilon \qquad \sigma' = \sigma \cup \{\ell.\mathsf{f} \mapsto \sigma(\ell.\mathsf{f}); \mathsf{e}\}$$

By T-Publish, $\Gamma \mid \Sigma \vdash \ell.\mathsf{f}.\mathsf{publish}(\mathsf{e}_0)$ : Unit and $\Gamma \mid \Sigma \vdash \mathsf{e}_0$ : Unit. It is easy to show that $\Gamma \mid \Sigma \vdash \sigma'$. Then, T-Empty finishes the case.

**Case** R-Let:

$$\mathsf{e} = \mathsf{let}\ \mathsf{x}{=}\ell\ \mathsf{in}\ \mathsf{e}_0 \qquad \mathsf{e}' = [\ell/\mathsf{x}]\mathsf{e}_0$$

By T-Let, $\Gamma \mid \Sigma \vdash \ell : \mathsf{E}$, $\Gamma, \mathsf{x} : \mathsf{D}' \mid \Sigma \vdash \mathsf{e}_0 : \mathsf{T}$, and $\mathsf{E} <: \mathsf{D}'$. By Lemma 1, $\Gamma \mid \Sigma \vdash [\ell/\mathsf{x}]\mathsf{e}_0 : \mathsf{T}'$ for some $\mathsf{T}'$ such that $\mathsf{T}' <: \mathsf{T}$. Letting $\mathsf{T}' = \mathsf{S}$ finishes the case.

Using the induction hypothesis, the cases for the congruence rules are easy. We show only the case that e is a field access.

**Case** Congruence (field access)

$$\mathsf{e} = \mathsf{e}_0.\mathsf{f} \qquad \mathsf{e}' = \mathsf{e}_0'.\mathsf{f}$$

By the induction hypothesis, $\Gamma \mid \Sigma \vdash \mathsf{e}_0 : \mathsf{C}_0$, $\Gamma \mid \Sigma \vdash \mathsf{e}_0' : \mathsf{D}_0$, $\mathsf{D}_0 <: \mathsf{C}_0$. By T-Field, $\Gamma \mid \Sigma \vdash \mathsf{e}_0.\mathsf{f} : \mathsf{C}$, $\Gamma \mid \Sigma \vdash \mathsf{e}_0 : \mathsf{C}_0$, *ftype*$(\mathsf{C}_0, \mathsf{f}) = \mathsf{n}\ \mathsf{C}$ for some n. It is easy to show that *ftype*$(\mathsf{C}_0, \mathsf{f}) =$ *ftype*$(\mathsf{D}_0, \mathsf{f})$. By T-Field, $\Gamma \mid \Sigma \vdash \mathsf{e}_0'.\mathsf{f} : \mathsf{C}$. □

The following lemma is required for the proof of Theorem 2.

**Lemma 3.** *If* $\Gamma \mid \Sigma \vdash e_0.m(\overline{e}) : C$, $\Gamma \mid \Sigma \vdash e_0 : C_0$, *and mtype*$(m, C_0) = \overline{D} \to T$, *then there exist* $\overline{x}$ *and* $e_0$ *such that mbody*$(m, C_0) = \overline{x}.e_0$ *and the lengths of* $\overline{x}$ *and* $\overline{D}$ *are equal.*

*Proof.* By induction on *mtype*$(\mathsf{m}, \mathsf{C}_0) = \overline{\mathsf{D}} \to \mathsf{T}$. □

We show the proof of Theorem 2 as follows.

*Proof.* By induction on $\emptyset \mid \Sigma \vdash \mathsf{e} : \mathsf{T}$.

**Case** T-Var: Cannot happen.

**Case** T-Field: $\mathsf{e} = \mathsf{e}_0.\mathsf{f}$

By the induction hypothesis, $\mathsf{e}_0$ is a location $\ell$ or $\sigma \mid \mu \mid \mathsf{e}_0 \longrightarrow \sigma' \mid \mu' \mid \mathsf{e}_0'$. In the latter case, the congruence rule finishes the case. In the former case, by T-Field, $\Gamma \mid \Sigma \vdash \ell : \mathsf{C}_0$. By T-Loc, $\Sigma(\ell) = \mathsf{C}_0$. As $\emptyset \mid \Sigma \vdash \mu$, $\mu(\ell) = \mathsf{new}\ \mathsf{C}_0(\ldots)$. By T-Field and the definition of *ftype*, *source*$(\mathsf{C}_0) = \overline{\mathsf{n}\ \mathsf{C}\ \mathsf{f}}$ and $\mathsf{f} \in \overline{\mathsf{f}}$, or *composite*$(\mathsf{C}_0) = \overline{\mathsf{n}\ \mathsf{C}\ \mathsf{f}{=}\mathsf{e}}$ and $\mathsf{f} \in \overline{\mathsf{f}}$. Thus, R-Field or R-FieldS finishes the case, respectively.

**Case** T-Invk: $\mathsf{e} = \mathsf{e}_0.\mathsf{m}(\overline{\mathsf{e}})$

By the induction hypothesis, there exist $i \geq 0$ and $\mathsf{e}_i$ such that all $\mathsf{e}_i$ are locations $\ell_0, \overline{\ell}$ or $\sigma \mid \mu \mid \mathsf{e}_i \longrightarrow \sigma' \mid \mu' \mid \mathsf{e}_i'$. In the latter case, the congruence rule finishes the case. In the former case, by T-Invk, $\Gamma \mid \Sigma \vdash \ell_0 : \mathsf{C}_0$. By T-Loc, $\Sigma(\ell_0) = \mathsf{C}_0$. As $\emptyset \mid \Sigma \vdash \mu$, $\mu(\ell_0) = \mathsf{new}\ \mathsf{C}_0(\ldots)$. By T-Invk and Lemma 3, *mbody*$(\mathsf{m}, \mathsf{C}_0) = \overline{\mathsf{x}}.\mathsf{d}$ for some $\overline{\mathsf{x}}$ and d such that the lengths of $\overline{\ell}$ and $\overline{\mathsf{x}}$ are equal. Then, R-Invk finishes the case.

**Case** T-New: $\mathsf{e} = \mathsf{new}\ \mathsf{C}(\overline{\mathsf{e}})$

By the induction hypothesis, there exist $i \geq 1$ and $\mathsf{e}_i$ such that all $\mathsf{e}_i$ are locations $\overline{\ell}$ or $\sigma \mid \mu \mid \mathsf{e}_i \longrightarrow \sigma' \mid \mu' \mid \mathsf{e}_i'$. In the latter case, the congruence rule finishes the case. In the former case, letting $\Sigma' = \Sigma \cup \{\ell \mapsto \mathsf{C}\}$ and R-New finish the case.

**Case** T-Assign: $\mathsf{e} = \mathsf{e}_0.\mathsf{f}_i = \mathsf{e}_1$

By the induction hypothesis, there exist $i = 0$ or $1$ and $\mathsf{e}_i$ such that all $\mathsf{e}_i$ are locations $\ell_0, \ell_1$ or $\sigma \mid \mu \mid \mathsf{e}_i \longrightarrow \sigma' \mid \mu' \mid \mathsf{e}'_i$. In the latter case, the congruence rule finishes the case. In the former case, by T-Assign, $\Gamma \mid \Sigma \vdash \ell_0 : \mathsf{C}_0$ and $source(\mathsf{C}_0) = \overline{\mathsf{n\ C\ f}}$. By T-Loc, $\Sigma(\ell_0) = \mathsf{C}_0$. As $\emptyset \mid \Sigma \vdash \mu$, $\mu(\ell_0) = \mathsf{new\ C}_0(\bar{\ell})$. It is easy to show that $\Gamma \mid \Sigma \vdash \mu \cup \{\ell_0 \mapsto [\ell_1/\ell_i]\mathsf{new\ C}_0(\bar{\ell})\}$. Then, R-Assign or R-AssignS finishes the case.
**Case** T-AssignCont: $\mathsf{e} = \{\mathsf{e}_0\}_{\ell.\mathsf{f}}$
By the induction hypothesis, $\mathsf{e}_0$ is an empty expression or $\sigma \mid \mu \mid \mathsf{e}_0 \longrightarrow \sigma' \mid \mu' \mid \mathsf{e}'_0$ (as $\Gamma \mid \Sigma \vdash \mathsf{e}_0 : \mathsf{Unit}$, $\mathsf{e}_0$ cannot be a location). In the latter case, the congruence rule finishes the case. In the former case, R-AssignCont finishes the case.
**Case** T-Cat: Easy.
**Case** T-Publish: $\mathsf{e}_0.\mathsf{f.publish(d)}$
By the induction hypothesis, $\mathsf{e}_0$ is a location $\ell_0$ or $\sigma \mid \mu \mid \mathsf{e}_0 \longrightarrow \sigma' \mid \mu' \mid \mathsf{e}'_0$. In the latter case, the congruence rule finishes the case. In the former case, R-Publish finishes the case.
**Case** T-Let: $\mathsf{let\ x{=}e_1\ in\ e_2}$
By the induction hypothesis, $\mathsf{e}_1$ is a location $\ell$ or $\sigma \mid \mu \mid \mathsf{e}_1 \longrightarrow \sigma' \mid \mu' \mid \mathsf{e}'_1$. In the latter case, the congruence rule finishes the case. In the former case, R-Let finishes the case.
**Cases** T-Loc and T-Empty: Immediate. □

We show the proof of Theorem 4 as follows.

*Proof.* As $\{\sigma'(\ell.\mathsf{g})\}_{\ell.\mathsf{g}}$ does not contain $\ell_0.\mathsf{f.subscribe(e_0)}$ as a subexpression, it is easy to show that there is a reduction $\sigma'' \mid \mu'' \mid \ell_0.\mathsf{f.subscribe(e_0)} \longrightarrow \sigma'' \uplus \{\ell_0.\mathsf{f} \mapsto \sigma''(\ell_0.f);\mathsf{e}_0\} \mid \mu'' \mid \mathsf{e}$ in the reduction closure $\longrightarrow^*$ where $\sigma''(\ell_0.\mathsf{f}) \supseteq \sigma(\ell_0.\mathsf{f})$. By the fact that $\ell_0.\mathsf{f} \in effect(\mu', \ell.\mathsf{g})$ and the definition of *handlers*, we have $\sigma'(\ell_0.\mathsf{f}) = \bar{\mathsf{e}}$. As $\sigma'(\ell_0.\mathsf{f}) \supseteq \sigma'' \uplus \{\ell_0.\mathsf{f} \mapsto \sigma''(\ell_0.f);\mathsf{e}_0\}$, we have $\mathsf{e}_0 \in \bar{\mathsf{e}}$. □

## About the authors

**Tetsuo Kamina** is an Associate Professor at Faculty of Science and Engineering, Oita University. He received his B.A. from International Christian University in 1999 and his M.A. and Ph.D. from the University of Tokyo in 2002 and 2005, respectively. His research interests are programming languages and software engineering, particularly modularization mechanisms and their applications to software development. Contact him at kamina@acm.org.

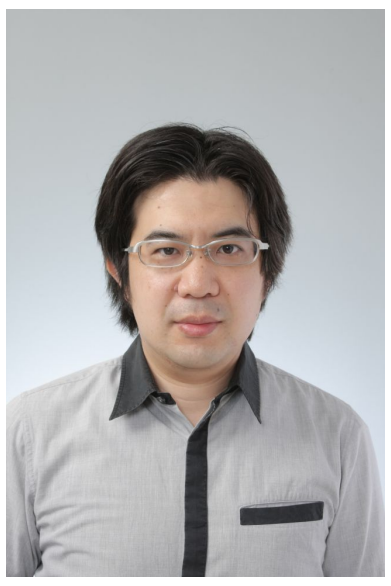

**Tomoyuki Aotani** is an assistant professor in Dept. of Mathematical and Computing Sciences, Tokyo Institute of Technology. He received his B.Sc. from Hosei University in 2004 and his M.A. and Ph.D. from the University of Tokyo in 2006 and 2009, respectively. His research interests include design and implementation of programming languages, program analysis and verification. Contact him at aotani@is.titech.ac.jp.

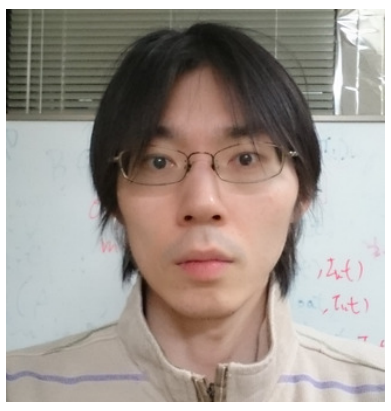